\documentclass[12pt]{article}\usepackage[]{graphicx}\usepackage[]{color}
\makeatletter
\def\maxwidth{ %
  \ifdim\Gin@nat@width>\linewidth
    \linewidth
  \else
    \Gin@nat@width
  \fi
}
\makeatother

\definecolor{fgcolor}{rgb}{0.345, 0.345, 0.345}

\usepackage{framed}
\makeatletter
 {\par\unskip\endMakeFramed%
 \at@end@of@kframe}
\makeatother

\definecolor{shadecolor}{rgb}{.97, .97, .97}
\definecolor{messagecolor}{rgb}{0, 0, 0}
\definecolor{warningcolor}{rgb}{1, 0, 1}
\definecolor{errorcolor}{rgb}{1, 0, 0}

\usepackage{alltt}
\usepackage[margin=1in]{geometry}
\usepackage[nodisplayskipstretch]{setspace}

\usepackage{tikz}
\usetikzlibrary{shapes,arrows}
\usepackage{multirow}

\usepackage{natbib}
\usepackage{amscd,amssymb,amsmath,mathrsfs,amsfonts,graphicx,verbatim,epsfig,
psfrag, float, fancybox} 
\usepackage{graphics}                 
\usepackage{graphicx}
\usepackage{color}                    
\usepackage{tikz}
\usetikzlibrary{shapes,arrows}

\usepackage{amsmath}    
\usepackage{amscd,amssymb,mathrsfs,amsfonts}
\usepackage{upgreek}
\usepackage{graphicx}   
\usepackage{verbatim}   
\usepackage{amsmath}
\usepackage{verbatim}
\usepackage{setspace} 
\usepackage{algorithmicx}
\usepackage{algorithm}
\usepackage{algpseudocode}

\usepackage{authblk}

\newcommand{\CTI}{\ensuremath{C^{\text{TI }}}}
\newcommand{\CTN}{\ensuremath{C^{\text{TN }}}}
\newcommand{\CNI}{\ensuremath{C^{\text{NI }}}}
\newcommand{\CNN}{\ensuremath{C^{\text{NN }}}}
\newcommand{\CI}{\ensuremath{\mathbf{C^{\text{I }}}}}
\newcommand{\CT}{\ensuremath{\mathbf{C^{\text{T }}}}}
\newcommand{\RR}{\ensuremath{\mathbf{R}}}
\newcommand{\CTE}{\ensuremath{\mathbf{C^{\text{T, eval }}}}}

\newcommand{\cT}{\ensuremath{\mathbf{c^{\text{T}}}}}
\newcommand{\rr}{\ensuremath{\mathbf{r }}}
\newcommand{\cTE}{\ensuremath{c^{\text{T, eval}}}}

\usepackage{subfig}
\usepackage{tabularx}
\usepackage{xcolor}
\usepackage{setspace}

\newcommand{\expit}{\text{expit}}

\DeclareMathOperator*{\argmin}{arg\,min}

\newcommand{\LL}{\mathcal{L}}

\newcommand{\YY}{\mathcal{Y}}

\newcommand{\FF}{\mathcal{F}}
\newcommand{\act}{\text{actual}}

\newcommand{\DO}{\text{D1}}
\newcommand{\DT}{\text{D2}}
\newcommand{\theo}{\text{theoretical}}

\newcommand{\unif}{\text{Uniform}}
\newcommand{\bernoulli}{\text{Bernoulli}}
\newcommand{\normal}{\text{Normal}}

\IfFileExists{upquote.sty}{\usepackage{upquote}}{}
\begin{document}
\title{Using Propensity Scores to Develop and Evaluate Treatment Rules with Observational Data}
\author{Jeremy Roth and Noah Simon}
\date{}
\maketitle
\begin{abstract}
In this paper, we outline a principled approach to estimate an individualized treatment rule that is appropriate for data from observational studies where, in addition to treatment assignment not being independent of individual characteristics, some characteristics may affect treatment assignment in the current study but not be available in future clinical settings where the estimated rule would be applied. The estimation framework is quite flexible and accommodates any prediction method that uses observation weights, where the observation weights themselves are a ratio of two flexibly estimated propensity scores. We also discuss how to obtain a trustworthy estimate of the rule's population benefit based on simple propensity-score-based estimators of average treatment effect. We implement our approach in the \texttt{R} package \texttt{DevTreatRules} and share the code needed to reproduce our results on GitHub.  
\end{abstract}
\doublespacing

\section{Introduction} \label{obs_sec:introduction}
Precision medicine strives to leverage an individual's specific characteristics to determine the most beneficial course of treatment for that individual. 
In this paper, we consider the practice of precision medicine in settings that present two particular sets of challenges: 1) data come from observational studies where individual characteristics may influence treatment assignment; and 2) the data at hand may measure individual characteristics that will not be available in future clinical settings. As we summarize in Table \ref{tab:selected_methods}, there are statistical methods to account for the observational study design in 1) but they do not account for the subtlety regarding variable roles in 2), and they often lack a user-friendly software implementation that would allow practitioners to reliably apply them.

Some characteristics that affect treatment recommendation in a particular study may be unavailable in future clinical settings, while other characteristics that did not directly affect treatment assignment may nonetheless be informative about treatment response in future clinical settings. For example, some information that influences treatment assignment (e.g. prognosis) may be measured subjectively in a given observational study and not be measured in that same manner in clinical settings outside the scope of the study; such a variable directly affects treatment assignment in the observational study but cannot directly influence treatment recommendations in future clinical settings. On the other hand, some available characteristics may not be interpretable in a clinical setting if there is a lack of scientific knowledge about their role in a disease pathway (e.g. gene expression levels); such characteristics do not directly affect the treatment decisions of clinicians but nonetheless may be predictive of treatment benefit. Each of these variable types should be handled differently by statistical methods that require distinct prediction of treatment assignment and prediction of outcome, but past statistical methods have not pointed out the distinction.

We propose a principled framework (along with a user-friendly implementation in the \texttt{R} package \texttt{DevTreatRules}, available on CRAN) that appropriately handles these distinct variable when developing and evaluating a \emph{treatment rule} based on data from an observational study where treatment is not independent of individual characteristics. A treatment rule is a function that recommends treatment based on individual characteristics; to be useful to practitioners, a treatment rule must appropriately handle individual characteristics as they are actually observed in clinical settings. Our framework yields a treatment rule that accounts for the clinically distinct roles of individual characteristics in an interpretable way by asking two simple questions -- 1) Which characteristics will be measured in the same manner in future settings where the estimated rule would be applied? 2) Which characteristics potentially influence treatment assignment and thus must be accounted for as confounders? -- that reflect clinical rather than statistical expertise. Our framework also places a clear emphasis on developing and evaluating the treatment rule on independent datasets which is also absent from other papers in the treatment rule literature, as discussed in Section \ref{obs_sec:gaps_in_literature}.


\section{Previous Work} \label{obs_sec:previous_work}
The framework presented in this paper draws on two previous bodies of work: 1) literature on estimating treatment rules;
and 2) tools from the causal inference literature that describe how to estimate the average treatment effect (ATE) in study designs where treatment is not randomized. There is notable overlap between these two topics even if the connection is not always made explicit in the literature; nearly all methods for estimating treatment effect that are appropriate in settings with non-randomized treatment rely on the inverse-probability-of-treatment weighting (IPW) approach to balance observed confounders across the treatment groups. The propensity score (the probability of treatment, conditional on individual characteristics) has played a vital role in facilitating reliable comparisons of outcomes across treatment groups in non-randomized studies at least back to \cite{Rosenbaum:1983}. \cite{Austin:2011} presents an excellent and accessible overview of the ATE derived from the causal inference literature and how it informs propensity-score-based estimation strategies for non-randomized treatment assignment (e.g. the IPW method).

In this section, we highlight previous work on estimating treatment rules and we defer discussion of the causal inference literature to Section \ref{obs_subsec:method_estimating}, where we can more clearly show how it shapes the target parameter of interest that is critical to our framework.

There is active statistical research on methods to predict the most beneficial treatment option for a specific individual, in line with the objectives of precision medicine. An existing approach generally belongs to one of two categories, either \emph{indirect} or \emph{direct}.

Indirect approaches (e.g. \cite{Kang:2014, Cai:2011, Lu:2013, McKeague:2014, Ciarleglio:2015}) typically assume structure on the regression function linking the conditional mean outcome to individual outcomes and a treatment indicator. The ideal treatment assignment for a individual is then inferred by predicting his or her expected outcomes across possible values of the treatment variable using the regression model, and choosing the treatment option with the most desirable predicted outcome (e.g. a larger mean time until relapse or a smaller probability of 5-year relapse). In \cite{Lipkovich:2017}, indirect approaches are given the label ``global outcome modeling''. As detailed in Section \ref{obs_sec:method}, our framework is an indirect approach.

On the other hand, direct methods are motivated by optimizing performance of the treatment rule itself in a population of interest rather then optimizing the accuracy of predicted outcomes and then using each individual-level predicted outcome to decide which treatment to assign (which would be the case with an indirect method). That is, these methods \emph{directly} estimate an optimal treatment rule rather than prioritizing estimation of the expected outcome conditional on individual characteristics and treatment assignment (i.e. a regression function), and then taking the additional step of assigning an individual to the treatment with the most desirable predicted outcome.

Direct methods generally seek the treatment rule within a particular class of allowable rules that maximizes an estimate of clinical benefit, and have the appealing motivation of not being vulnerable to mis-specification of the regression function predicting outcome based on individual characteristics (though direct methods must still make other modeling assumptions to which they are sensitive). \cite{Lipkovich:2017} present an outstanding recent survey of statistical methods for estimating treatment rules in RCTs; direct methods are part of that review's ``optimal treatment regimes'' category. \cite{Zhang:2015} pre-specify that the treatment rule must be a nested sequence of ``if, then'' statements with one or two individual characteristics involved in each statement. \cite{Zhang_robust:2012} take a similar approach but instead use a regression model to dictate the rule's structure.

Other direct methods make less restrictive assumptions about the form of the treatment rule at the cost of interpretability of the rule. \cite{Zhao:2012} propose outcome-weighted learning (OWL), which defines the optimal treatment rule as the solution to a weighted classification problem whose solution can be approximated using a modified form of support vector machines (SVM), a well-established statistical learning tool \citep{Hastie:2008}. As noted by \cite{Zhang:2012}, OWL can be viewed as part of a general weighted-classification framework for estimating a treatment rule, so any classification method that accommodates observation weights (e.g. classification trees or penalized regression \citep{Hastie:2008}) is a viable alternative to SVM in the estimation stage of OWL. The interactions-based procedure from \cite{Chen:2017} based on the earlier work of \cite{Tian:2014} is a recent example of a promising direct approach to estimating treatment rules with even greater flexibility.

Importantly, as mentioned in \cite{Lipkovich:2017}, some existing direct and indirect methods for estimating treatment rules are adaptable to observational study designs where treatment assignment is not independent of individual characteristics by accommodating the IPW approach \citep{Austin:2011}, which re-weights observations by the inverse of their estimated propensity scores so clinically observed confounders are roughly balanced between the treatment groups. The IPW adjustment thus allows for a sensible direct comparison of mean re-weighted outcomes across treatment groups that is reflective of the underlying population where the rule would be applied in the future. There is no consensus among researchers on whether the direct or indirect approach to estimating treatment rules yields superior results.

\section{Categorization of Individual Characteristics}
We propose partitioning each individual characteristic collected in an observational study (aside from outcome and treatment variables) into the four clinically distinct categories shown in Table \ref{tab:variable_roles}: \emph{\CTI}, \emph{\CTN}, \emph{\CNI}, or \emph{\CNN}, where the abbreviations in each superscript tell us whether each characteristic (C) affects treatment assignment in the current study (TN), is expected to be observed in independent studies (NI), both (TI), or neither (NN). This helps make explicit (as is not done in previous work) that only the \CTI and \CNI variables are viable candidates for inclusion in a treatment rule because the remaining \CTN and \CNN will not be available in future clinical situations where a proposed treatment rule would be implemented. Here is a brief example of how each variable type may present itself in a hypothetical observational study:
\begin{enumerate}
  \item \textbf{Example of \CTI: Age.} In the study population, physicians might have been more likely to recommend treatment to older individuals. In independent clinical settings, we are confident that age can be reliably measured on the same scale.
  \item \textbf{Example of \CTN: Center-specific measure of prognosis.} In the study population, clinicians may have been more likely to recommend treatment to individuals with a poor prognosis as estimated by a center-specific set of guidelines (e.g. a hospital's standard rule-of-thumb procedure based on their specific doctors' prior experiences with individuals from the population). In independent clinical settings taking place in different centers, clinicians would not estimate an individual's prognosis with that same center-specific approach.
  \item \textbf{Example of \CNI: Gene expression levels.} In the study population, individuals' gene expression levels may have been measured but clinicians did not consider the information when recommending treatment due to a lack of scientific knowledge about the genes' roles in disease progression and response to treatment. In independent clinical settings, gene expression levels might still be reliably measured and would be eligible to inform treatment decisions if a newly developed treatment rule suggests their importance or if other scientific knowledge becomes available in the interim.
  \item \textbf{Discussion of \CNN:} \CNN consists of variables in a dataset that are believed to have no role in influencing treatment and cannot be reliably collected in future clinical settings; the variables in \CNN are not of interest to development or evaluation of treatment rules. An example could be a study ID variable.
\end{enumerate}
It will be useful in later sections to define: $\CI \equiv (\CTI, \CNI)$ as all observed individual characteristics that are also expected to be observed in independent clinical settings. Also, we define $\RR$ as a subset of the variables contained in $\CI$ that the researcher believes may affect response to treatment and thus are viable candidates for a treatment rule. 
We also define $\CT \equiv (\CTI, \CTN)$ as all individual characteristics that potentially affect treatment in the current observational study. In the \texttt{BuildRule()} and \texttt{EvaluateRule()} functions from \texttt{DevTreatRules}, users are required to provide the characteristics in $\CT$ with the argument \texttt{names.influencing.treatment} and the characteristics in $\RR$ with the argument \texttt{names.influencing.rule}. That is, individual characteristics are never entered into the package without first being categorized by the user.

\section{Gaps in the Literature}\label{obs_sec:gaps_in_literature}
To help situate this paper in the literature, Table \ref{tab:selected_methods} presents a selection of direct and indirect methods for estimating treatment rules and whether each satisfies five criteria. As seen in Table \ref{tab:selected_methods}, none of these selected previous methods (and, to our knowledge, no other method in the literature) distinguishes between observed individual characteristics using the clinically meaningful categorization of whether they will be available in future clinical settings (in which case they are sensible candidates for building the treatment rule) or are not expected to be available in future settings (in which case they should not be used to build the rule). 
The lack of available R packages implementing previous approaches is also re-enforced by Table \ref{tab:selected_methods}.

We believe that our work fills a gap in the literature by providing practitioners with a principled approach to appropriately classify variable types as in Table \ref{tab:variable_roles} and by providing the user-friendly \texttt{DevTreatRules}, which actually requires users to make their clinically informed variable categorizations when they apply the work to real data (using the arguments \texttt{names.influencing.treatment} and \texttt{names.influencing.rule} in its \texttt{BuildRule()} and \texttt{EvaluateRule()} functions). In addition, although it is not considered in the table, none of those previous methods explicitly integrates data-splitting to ensure that the stages of development/evaluation (or development/validation/evaluation) of a treatment rule are conducted on independent datasets to yield a trustworthy estimate of the rule's population impact; we emphasize data-splitting throughout this paper and in the formalized procedure in Section \ref{obs_subsec:recipe}.

\section{Motivating Example} \label{obs_sec:motivating_example}
\tikzstyle{block} = [rectangle, draw, fill=blue!20, 
    text width=5em, text centered, rounded corners, minimum height=4em]
\tikzstyle{block2} = [rectangle, draw, fill=orange!20, 
    text width=5em, text centered, rounded corners, minimum height=4em]
\tikzstyle{block3} = [rectangle, draw, fill=green!20, 
    text width=5em, text centered, rounded corners, minimum height=4em]
\tikzstyle{line} = [draw, -latex']
\tikzstyle{cloud} = [draw, ellipse,fill=red!20, node distance=3cm,
    minimum height=2em]

Suppose an observational study recruits individuals with a particular type of cancer from a single hospital and collects baseline information at the time of recruitment. Further suppose that, for each individual, this observational dataset measures: months until relapse ($Y$); an indicator of receiving standard-of-care ($T=0$) or additional chemotherapy in addition to the standard-of-care option ($T=1$); age; a measure of day-to-day life functioning; and expression levels of $p$ genes. For simplicity, we assume that clinicians decide to treat individuals based only on age and day-to-day life functioning. In contrast, the gene expression levels are unobservable by clinicians due to high cost and are also uninterpretable due to a lack of scientific knowledge about the genes' roles in disease progression and response to treatment. In this illustrative example, we assume no unmeasured confounding.\footnote{In practice, however, gene expression levels may serve as surrogates for unobserved confounders. For instance, it is estimated that 20\%-30\% of breast cancer cases consist of an over-expression of human epidermal growth factor receptor type 2 (HER2), which can be targeted by additional treatment that inhibits HER2 generation \citep{Joensuu:2006, Hudis:2007}. For types of cancer with unknown subtypes the gene expressions conducted in an observational study may be associated with (but not explicitly define, as with HER2) disease subtypes that may be more or less vulnerable to disruption by the treatment mechanism.}

In practice, there are two alternative scores used to an individual's day-to-day functioning. One is the Eastern Cooperative Oncology Group (ECOG) Performance Status, a grade ranging from 0 to 5 where a lower grade represents fewer restrictions in day-to-day life \citep{Oken:1982}. The other is the Karnofsky Performance Status (KPS), an 11-point scale ranging from 0 to 100 where a lower value represents more day-to-day restrictions \citep{Karnofsky:1949}. As an added complication, we suppose that in this particular observational study day-to-day functioning is measured using the ECOG score, but that KPS is used instead of ECOG score in future clinical settings where we would like to apply our developed treatment rule in the future.\footnote{Although in practice there are mappings between ECOG and KPS, for illustrative purposes here we treat them as variables representing the same individual characteristic using qualitatively distinct scales (e.g. higher scores mean lower quality of life for KPS but a higher quality of life for ECOG) and suppose they are non-conformable.} In contrast, we assume an individual's age and $p$ gene expression measurements will be reliably collected in future clinical settings.

\begin{figure}[H]
  \centering
  \begin{tikzpicture}[node distance = 2.9cm, transform shape]
    \node [block] (T) {Treatment};
    \node [block2, above of=T] (ECOG) {ECOG \\ (\CTN)};
    \node [block3, left of=ECOG] (Age) {Age \\ (\CTI)};
    \node [block2, right of=ECOG] (KPS) {KPS \\};
    \node [block, below of=T] (Y) {Years until relapse};
    \node [block3, left of=Y] (X) {Gene expression levels \\ (\CNI) };
    \path [line] (Age) to (Y);
    \path [line] (Age) to (T);
    \path [line] (ECOG) to (T);
    \path [line] (ECOG) to[out=325, in=45] (Y);
    \path [line] (ECOG) to (KPS);
    \path [line] (KPS) to (ECOG);
    \path [line] (KPS) to[out=270, in=0] (Y);
    \path [line] (T) to (Y);
    \path [line] (X) to (Y);
\end{tikzpicture}
\caption{Potential data-generating mechanism. Each node shows a individual characteristic (and, for those besides treatment and outcome, its corresponding variable type from Table \ref{tab:variable_roles} in parentheses)}
\label{fig:motivating_example_mechanism}
\end{figure}
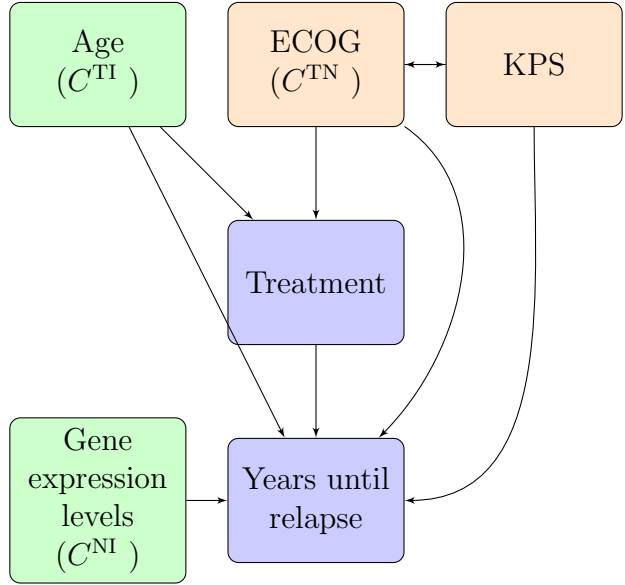

Figure \ref{fig:motivating_example_mechanism} shows our hypothesized data-generating mechanism. We have the following variable types in this example: \CNI $=$ (gene$_1$, \ldots, gene$_p$), \CTI $=$ age, \CTN $=$ ECOG, \CT $=$ (age, ECOG), and \RR$=$(gene$_1$, \ldots, gene$_p$). The KPS variable is a potential confounder only in independent clinical settings (but not the current dataset) and as such KPS plays an important role in evaluation of the rule that we discuss in Section \ref{obs_subsec:method_evaluating}.

We are interested in building and evaluating a classifier, for future individuals, that indicates their optimal treatment based on their gene expression values. However, in constructing this classifier, we must account for a) the confounding influence of age and day-to-day functioning on the relationship between treatment and months until relapse; and b) the distinct approaches to measuring day-to-day functioning across different clinical settings (based on use of ECOG or KPS). 

\section{Method} \label{obs_sec:method}
Here, we provide some theoretical justification for how the individual characteristics \CTN, \CTI, and \CNI  should be used to develop a treatment rule and evaluate the rule's benefit. We will refer to this method as the \emph{split-regression} approach to developing a treatment rule. As in Section \ref{obs_sec:motivating_example}, we will interpret $T$ as an indicator of receiving a new treatment in addition to standard-of-care ($T=1$) or standard-of-care alone ($T=0$) and outcome $Y$ as months until relapse.

\subsection{Estimating the Rule} \label{obs_subsec:method_estimating}
\subsubsection{The Target Parameter} \label{obs_subsubsec:method_estimation_target}
Our goals are 1) to identify a subset of the population -- in terms of $\RR$, the researcher-chosen subset of individual characteristics that are valid candidates for inclusion in a treatment rule -- we expect to benefit from treatment and 2) to estimate the extent of benefit in this subpopulation. 

Our estimation strategy begins with a foundational parameter of interest, $E[Y^{1} - Y^{0}]$, often called the \emph{average treatment effect} (ATE) in the causal inference literature, 
where in \emph{potential outcomes} notation, $Y^1$ is the months until relapse that would have been observed had an individual received treatment and $Y^0$ is the months until relapse that would have been observed had the individual received standard-of-care (see \cite{Kennedy:2015} for an excellent review). So the ATE gives the average difference in months until relapse when an individual in the population of interest receives treatment instead of standard-of-care. The ATE is identifiable under three assumptions: 1) \emph{consistency}: $T=t$ implies $Y=Y^t$; 2) \emph{no unmeasured confounding}: conditional on observing $\CT$, $T$ is independent of $Y^t$; 3) \emph{positivity}: if $P(\CT > 0)$ then $P(T=t \mid \CT=\cT)>0$, for $t=0, 1$.

As originally developed by \cite{Robins:1986},  if the consistency, no unmeasured confounding, and positivity assumptions hold, then the ATE is equivalent to
\begin{equation}
    \psi \equiv \int_\CT \left\{E\left[Y \mid \CT, T=1\right] - E\left[Y \mid \CT, T=0 \right]\right\}dP(\CT) \label{eq:rewritten_ATE},
\end{equation}
which is known as a ``g-computation'' formula in the causal inference literature. To estimate whether treatment offers a superior outcome for an individual with characteristics $\RR=\rr$, we simply consider the subgroup-specific ATE
\begin{equation}
  E[Y^{1} - Y^{0} \mid \RR=\rr] \label{eq:conditioned_ATE}.
\end{equation}
The subgroup-specific ATE in \eqref{eq:conditioned_ATE} is also a parameter of interest in \cite{Cai:2011}, the previous method that we believe is most similar to the one we discuss in this section. The practical limitations of \cite{Cai:2011} on which this paper expands are that it was designed only for the RCT setting, did not explicitly distinguish between individual characteristics $\CT$ and $\CI$, was not implemented in an R package, and did not share code to guide users in implementation.

To estimate \eqref{eq:conditioned_ATE} in the setting of non-randomized treatment assignment, the g-computation formula in \eqref{eq:target_ATE} similarly changes to
\begin{equation}
  \resizebox{.90 \textwidth}{!} 
  {
    $\psi(\rr) \equiv \int_\CT \left\{E\left[Y \mid \CT, \RR=\rr,T=1\right] - E\left[Y \mid \CT, \RR=\rr, T=0 \right]\right\}dP(\CT | \RR=\rr)$ \label{eq:target_ATE},
    }
\end{equation}
which estimates the average treatment effect for individuals with characteristics $\RR=\rr$ that will be observable in future clinical settings.

From \eqref{eq:target_ATE}, we see that we should treat the individuals defined by $ \Omega^+ = \left\{\rr\mid \psi(\rr) > C \right\} \label{eq:population_to_treat}$ for $C \geq 0$. If $\RR$ were a set of gene expression levels and $C=0$, for example, then $\Omega^+$ would be the subset of gene expression levels for which treatment increases the expected number of months until relapse. The expected improvement in months until relapse among the treated subpopulation is $\int_{\rr\in\Omega^+} \psi(\rr) dP(\rr) \label{eq:population_treatment_benefit}.$

\subsubsection{Estimating the Target Parameter} \label{obs_subsubsec:method_estimation_target_estimating}
One possible approach for estimating the modified ATE in \eqref{eq:target_ATE} would be to separately estimate $E\left[Y \mid \CT, \RR, T=t\right]$ for $t=0, 1$ and estimate $dP(\CT | \RR )$, then plug these estimates into \eqref{eq:target_ATE}; however, this approach requires estimation of a conditional density function that is only practical when the individual characteristics in $\RR$ are perhaps one or two categorical variables with very few levels while in other situations the approach would prescribe a very complicated and highly variable average (as described in greater detail in Chapter 3 of \cite{Varadhan:2013}, for example).

Our goal is now to re-write our estimation target \eqref{eq:target_ATE} as a simple minimization problem that does not involve estimation of the conditional density $dP(\CT | \RR)$. We begin by re-stating \eqref{eq:target_ATE} as $\psi(\rr) = f_1(\rr) - f_0(\rr)$, where, for $t=0, 1$, 
\begin{equation}
  f_t(\rr) = \int_\CT E\left[Y\mid \CT, \RR=\rr,T=t\right]dP(\CT | \RR=\rr). \label{eq:target_rewritten_one_group}
\end{equation}
We emphasize that \eqref{eq:target_rewritten_one_group} is exactly equivalent to \eqref{eq:target_ATE}, just with altered notation. By taking the perspective in \eqref{eq:target_rewritten_one_group} we only need to estimate $f_t(\rr)$ for $t \in \{0, 1\}$ to obtain an estimate of the $\psi(\rr)$ in \eqref{eq:target_ATE}. As derived in the Appendix, it turns out that $f_t(\rr)$ can be written as the minimizer
\begin{equation}
  f_t(\rr) \equiv \argmin_{f\in\mathcal{F}} \int_\RR \int_\CT w_t(\rr, \cT) \LL(y, f(\rr)) dP(y, \cT,\rr | T=t), \label{eq:weighted_target}
\end{equation}
where  $\LL(y, f(\rr))$ is any function of y and \rr \: such that its conditional expectation $E[y \mid f(\rr)]$ is the minimizer -- which in practice we may think of as a ``canonical'' loss function such as squared-error loss for a continuous $y$ or logistic loss for a binary $y$  -- and $\mathcal{F}$ is a function class in which the rule is known to lie (one possibility could be $\ell_1(P)$, the space of all absolutely integrable functions over $P$). 

A natural weight function $w_t(\rr, l)$ that can be used turns out to be
\begin{equation}
  w_t(\rr, \cT) = \frac{P(T=t | \RR=\rr)}{P(T=t | \RR=\rr, \CT=\cT)} \label{eq:weight_function}.
\end{equation}
We note that the standard IPW observation weight would replace the numerator of \eqref{eq:weight_function} with a 1. 
In fact, the weight function in \eqref{eq:weight_function} is closely related to the ``stabilized weights'' proposed by \cite{Robins:2000}, which in this case would be $P(T=t) / P(T=t | \CT=\cT, \RR=\rr)$. The weight function implied by \cite{Robins:2000} differs from \eqref{eq:weight_function} by the latter's extra conditioning on $\RR=\rr$, the subset of individual characteristics that are potentially informative inputs to the treatment rule. 

The derivation in \eqref {eq:weighted_target} implies a natural estimate of $f_t$ that is not complicated by the dimensionality of $\RR$: the minimizer of the weighted sample average over individuals in the $T=t$ group
\begin{equation}
  \tilde{f}_t \equiv \operatorname{argmin}_{f\in\mathcal{F}} \frac{1}{n_t} \sum_{T_i = t} \left( \frac{\tilde{P}(T=t \mid \rr_i)}{\tilde{P}(T=t \mid \rr_i,\cT_i)}\right) \LL(y_i, f(\rr_i)) \label{eq:sample_average}, 
\end{equation}
where $n_t$ is the number of individuals in the group $T=t$, $\tilde{P}(T=t \mid \rr_i)$ is an estimate of $P(T=t \mid \RR=\rr)$, and $\tilde{P}(T=t | \rr_i, \cT_i)$ is an estimate of $P(T=t \mid \RR=\rr_i, \CT=\cT_i)$. We note that the estimate $\tilde{f}_t$ is only reasonable if $\mathcal{F}$ is a suitably constrained class (e.g. a class with smoothness constraints, like a Sobolev class or class with bounded total variation \citep{VanDeGeer:2000}).

However, the formula~\eqref{eq:sample_average} suggests another approach for estimation of $f_t$: Instead of necessarily solving a formal empirical minimization as in \eqref{eq:sample_average}, one might use \emph{any} predictive modeling method that accommodates observation weights (e.g. generalized linear models, lasso, ridge regression, boosted trees, neural nets, and many others). All of these methods can be written as, either exactly or approximately, minimizing a weighted least-squares-like loss over a, potentially complicated, function class.  Many of the methods indicated in Table \ref{tab:selected_methods} as being flexible in fact only support penalized or non-penalized weighted regression; our framework supports much more flexibility by moving beyond regression-based methods.

Now that we can estimate $\tilde{\psi}(\rr) = \tilde{f}_1(\rr) - \tilde{f}_0(\rr), \label{eq:estimator_target_ATE}$ we can simply form the treatment rule as $\tilde{B}(\rr) \equiv I\left[\tilde{f}_1(\rr) - \tilde{f}_0(\rr) > 0 \right]$, which recommends treatment to an individual with characteristics $\RR=\rr$ if the estimated months until relapse is higher under $T=1$ than $T=0$. 

\subsection{The Recipe} \label{obs_subsec:recipe}
The work in Sections \ref{obs_subsec:method_estimating} yields the following procedure applied to a development dataset (D1), which must be independent of the evaluation dataset (D2); we use the superscripts D1 and D2 to emphasize the dataset on which the accompanying estimate is formed.

\begin{enumerate}
  \item \label{recipe:D1_definitions} Use the scientific knowledge underlying D1 to partition observed individual characteristics into the four categories presented in Table \ref{tab:variable_roles}: \CTI, \CTN, \CNI, and \CNN. Also form $\CT = (\CTI, \CTN)$, $\CI = (\CTI, \CNI)$, and form the potential inputs for the treatment rule $\RR \subseteq \CI$.
  \item \label{recipe:D1_prediction} For observations $i=1, \ldots, n$ on D1:
    \begin{enumerate}
     \item Choose a prediction method and estimate the propensity scores  $\tilde{P}^{\DO}(T=1 \mid \RR=\rr_i)$ and $\tilde{P}^{\DO}(T=1 \mid \RR=\rr_i, \CT=\cT_i)$.\footnote{It can be useful to truncate estimated propensity scores so they are not too close to 0 or 1 (which can lead to very large observation weights); a default setting in our software implementation truncates estimated propensity scores to stay between 0.05 and 0.95, but this choice can be overwritten by the user.}
     \item Compute the weights $\tilde{W}_t(\RR=\rr_i, \CT=\cT_i) = \frac{\tilde{P}^{\DO}(T=t \mid \RR=\rr_i)}{\tilde{P}^{\DO}(T=t \mid \RR=\rr_i, \CT=\cT_i)}$, for $t=0,1$.
     \item Choose a prediction method that accommodates observation weights (e.g. generalized linear regression, lasso, boosted trees, and many others) and estimate $\tilde{f}_0$ and $\tilde{f}_1$ as suggested by \eqref{eq:sample_average}. For example, weighted linear regression with a continuous response would yield, for observations $i=1, \ldots, n$ on D1,
    \begin{align}
      \tilde{\beta}_0^{\DO} &\equiv \argmin_{\beta \in \mathbb{R}^p} \frac{1}{n_0} \sum_{T_i = 0} \tilde{W}_0(\rr_i, \cT_i) (y_i - \rr_i^{\top}\beta)^2, \\
      \tilde{\beta}_1^{\DO} &\equiv \argmin_{\beta \in \mathbb{R}^p} \frac{1}{n_1} \sum_{T_i = 1} \tilde{W}_1(\rr_i, \cT_i) (y_i - \rr_i^{\top}\beta)^2,
    \end{align}
    where $\tilde{W}_t(\RR=\rr_i, \CT=\cT_i) = \frac{\tilde{P}^{\DO}(T=t \mid \RR=\rr_i)}{\tilde{P}^{\DO}(T=t \mid \RR=\rr_i, \CT=\cT_i)}$ for $t=0,1$ and where we define $\tilde{f}_0^{\DO}(\rr) \equiv \rr^{\top} \tilde{\beta}_0^{\DO} $ and  $\tilde{f}_1^{\DO}(\rr) \equiv \rr^{\top} \tilde{\beta}_1^{\DO} $.
    \end{enumerate}
\item \label{recipe:D1_form_rule} Form the treatment rule $\tilde{B}(\rr) \equiv I\left[\tilde{f}_1^{\DO}(\rr) - \tilde{f}_0^{\DO}(\rr) > 0\right]$, where $I(\cdot)$ is the indicator function.
\item \label{recipe:D2_definitions} Use scientific knowledge underlying D2 to select the potential confounders $\CTE$. \footnote{Ideally, D1 and D2 would be datasets from separate observational studies where D2 independently samples from the population where future intervention would take place, but D1/D2 may also be a random partition of data from a single observational study; in the latter case, we will have $\CTE = \CT$. The \texttt{DevTreatRules} package supports developing and evaluating rules in either situation.}
\item \label{recipe:D2_evaluate_rule} For observations $j=1, \ldots, m$ on D2:
  \begin{enumerate}
    \item Assign the recommended treatment with $\tilde{B}_j^{\DT} \equiv \tilde{B}^{\DT}(\rr_j)$.
    \item As discussed next in Section \ref{obs_subsec:method_evaluating}, form the IPW-based estimators of the ATE in the test-positives and in the test-negatives using \eqref{eq:estimate_ATE_IPW} and \eqref{eq:estimate_ATE_test_negatives}, respectively.
  \end{enumerate}
\end{enumerate}

\subsection{Evaluating the Rule} \label{obs_subsec:method_evaluating}
Now we define $\CTE$ as the set of potential confounders of the association between treatment and response \emph{in the evaluation dataset}. We note that if the development and evaluation datasets are partitions of the same observational dataset then the variables in $\CTE$ will be identical to those in $\CT$, but this need not be the case when development and evaluation are carried out using data from separate studies. For example, in the motivating example from Section \ref{obs_sec:motivating_example}, we had \CT$=$(age, ECOG) but \CTE $=$ (age, KPS).

To evaluate the the developed rule $\tilde{B}$ using Section \ref{obs_subsec:recipe}, we can use an estimate of the ATE in the test-positives population with the IPW-based estimator (see e.g. \cite{Austin:2011})
\begin{equation}
    \resizebox{.90 \textwidth}{!} 
  {
  $\widehat{\text{ATE}}^+ \equiv \frac{1}{N^+} \sum\limits_{\{ j \: \mid \: \tilde{B}(\rr_j) = 1 \}} \frac{t_j y_j}{\tilde{P}(T=1 \mid \CTE=\cTE)} - \frac{1}{N^+} \sum\limits_{\{ j \: \mid \: \tilde{B}(\rr_j) = 1 \}} \frac{(1-t_j) y_j}{\tilde{P}(T=0 \mid \CTE=\cTE)}$, \label{eq:estimate_ATE_IPW}
    }
\end{equation}
where $N^+$ is the number of test-positives. A small modification to \eqref{eq:estimate_ATE_IPW} also estimates the effect of avoiding treatment among the test-negatives:
\begin{equation}
    \resizebox{.90 \textwidth}{!} 
  {
  $ \widehat{\text{ATE}}^- \equiv \frac{1}{N^-} \sum\limits_{\{ j \: \mid \: \tilde{B}(\rr_j) = 0 \}} \frac{t_j y_j}{\tilde{P}(T=1 \mid \CTE=\cTE)} - \frac{1}{N^-} \sum\limits_{\{ j \: \mid \: \tilde{B}(\rr_j) = 0 \}} \frac{(1 - t_j) y_j}{\tilde{P}(T=0 \mid \CTE=\cTE)}$, \label{eq:estimate_ATE_test_negatives}
    }
\end{equation}
where $N^-$ is the number of test-negatives. We note that we would expect \eqref{eq:estimate_ATE_test_negatives} to be a negative number for a rule that accurately identifies the test-negatives.

We can also form a simple estimator of the average benefit of the rule (ABR) in the population from which our the evaluation dataset is a representative sample with 
\begin{equation*}
  \widehat{\text{ABR}} \equiv \left(\frac{N^+}{N^+ + N^-}\right) \widehat{\text{ATE}}^+ + \left(\frac{N^-}{N^+ + N^-}\right) \left(-\widehat{\text{ATE}}^-\right),
\end{equation*}
a weighted average of the benefit of receiving treatment the test-positives from \eqref{eq:estimate_ATE_IPW} and the benefit of \emph{avoiding} treatment in the test-negatives from \eqref{eq:estimate_ATE_test_negatives}, respectively.

Alternatively one could estimate ATE with the estimator developed by \cite{Robins:1994} that is sometimes called the \emph{doubly-robust} or \emph{augmented} analog of \eqref{eq:estimate_ATE_IPW}; \cite{Lunceford:2004} present an excellent derivation and simulation study comparing different estimators of the ATE.  We chose to present only the IPW-based estimators for ease of exposition.

\subsection{R Implementation} \label{obs_subsec:R_implementation}
The \texttt{R} package \texttt{DevTreatRules} implements this paper's split-regression approach. In particular, the functions \texttt{SplitData()}, \texttt{BuildRule()}, and \texttt{EvaluateRule()} handle, respectively: the development/evaluation partitioning of a dataset (or development/validation/evaluation partitioning if model selection is also performed); the development of the treatment rule (in dataset D1) as in steps 2-4 of Section \ref{obs_subsec:recipe}; and the evaluation of the rule (in dataset D2) as in steps 5-6 of Section \ref{obs_subsec:recipe}. The vignette accompanying \texttt{DevTreatRules} walks through an example of building and evaluating a treatment rule using the package, in a situation where model selection is also performed using the \texttt{CompareRulesOnValidation()} function.

\section{Simulations} \label{obs_sec:simulations}
We simulate data for a development sample of size $n$ with the individual characteristics
\begin{align*}
  X_i \sim \unif(0, 2), \hspace{10mm} L_i \sim \bernoulli(0.5), \hspace{10mm} G_ i \sim \normal(0, 1)
\end{align*}
the treatment indicator 
\begin{align*}
  T_i \mid L_i &\sim \left\{
     \begin{array}{lr}
       \bernoulli(0.75), & L_i=0\\
       \bernoulli(0.25), & L_i=1,
     \end{array}
   \right. 
\end{align*}
and binary outcome
\begin{align*}
  P(Y_i =1 \mid X_i, L_i, T_i) &= \left\{
  \begin{array}{lr}
       \expit \left[\beta_{0, T=0} + \beta_{1, T=0} X_i + \gamma_{T=0} L_i\right], &  T_i=0\\
       \expit \left[\beta_{0, T=1} + \beta_{1, T=1} X_i + \gamma_{T=1} L_i \right], &  T_i=1,
     \end{array}
   \right. 
\end{align*}
for $i=1, \ldots, n$, where $\beta_{0, T=t}, \beta_{1, T=t}, \gamma_{T=t} \in \mathbb{R}$ for $t=0,1$. Using the notation from Section \ref{obs_sec:introduction}, we would categorize $\CT=\CTE=L$ and $\CI=\RR=(X, G)$.

\begin{figure}[H]
  \centering
  \begin{tabular}{ccl} 
    \vspace{-8mm}
    \textbf{Sample averaging} & \textbf{Population averaging} \\
    \vspace{-2mm}
    \includegraphics[width=0.35\linewidth]{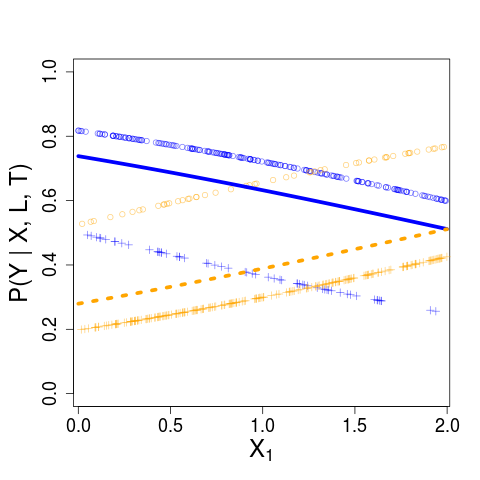} & \includegraphics[width=0.35\linewidth]{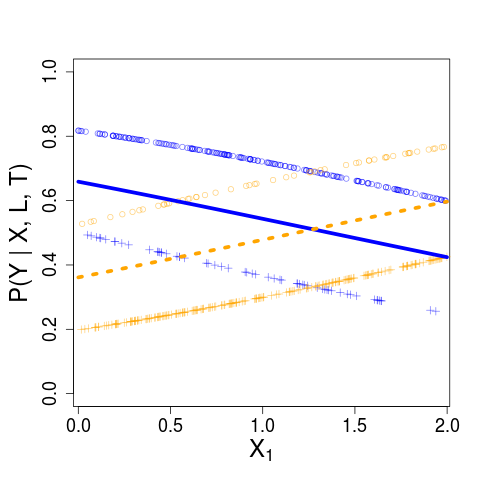} 
    & \hspace{-8mm} \includegraphics[width=0.4\linewidth]{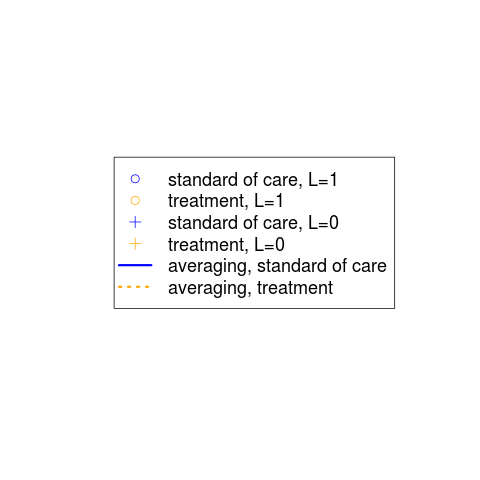}
  \end{tabular}
      \caption{Simulation scenario with $n=500$, $(\beta_{0, T=0}, \beta_{1, T=0}, \gamma_{T=0}) = (0, -0.55, 1.5)$, and $(\beta_{0, T=1}, \beta_{1, T=1}, \gamma_{T=1})=(-1.4, 0.55, 1.5)$}
    \label{fig:simulations_one_scenario}
\end{figure}

Figure \ref{fig:simulations_one_scenario} depicts the relationship between $P(Y \mid X, L, T)$ and $X$, where the true response probability for the $L=1$ group is shown with circles and for the $L=0$ group with triangles. The standard-of-care group is shown in blue and the treatment group is shown in orange. As seen in Figure \ref{fig:simulations_one_scenario}, the outcome ($Y$) is more likely under treatment than under standard-of-care for individuals with a value of $X$ about 1.3 in both the $L=1$ and $L=0$ groups. Thus, the optimal treatment rule would recommend treatment to an individuals with $X>1.3$. However, due to the confounding effect of $L$, an empirical average of the response-curves for each treatment group (solid blue line and dotted orange line in the left panel of Figure \ref{fig:simulations_one_scenario}) incorrectly suggests that there is no subset of individuals for whom treatment makes the outcome more likely. On the other hand, the IPW approach (solid blue line and dotted orange line in the right panel of Figure \ref{fig:simulations_one_scenario}) uses a weighted average of response-curves, where each weight is the IPW weight (based on $L$), to correctly identify the value of $X$ (about 1.3) where the response-curves cross.

In Table \ref{tab:simulation_results_mean_response}, we present the mean probability of the (desirable) outcome  for a range of development set sample sizes, specifying logistic regression (with $\RR$ as the predictors) for steps \ref{recipe:D1_prediction} in Section \ref{obs_subsec:recipe}. The first row shows the mean outcome probability for rules built using the split-regression approach described in Section \ref{obs_subsec:recipe}. The second row reports the mean outcome probability for a modification to the split-regression approach that ``naively''  uses the incorrect sample averaging shown in the left panel of Figure \ref{fig:simulations_one_scenario} (i.e. it uses identical observation weights in step \ref{recipe:D1_prediction} of Section \ref{obs_subsec:recipe}).

The estimated outcome probabilities in the first three rows of Table \ref{tab:simulation_results_mean_response} can be compared to the benchmark value of $0.574$ for the optimal rule (known from the data-generating mechanism) that perfectly assigns treatment to only those who benefit and withholds it from those who do not benefit. We also compare to a rule that recommends everyone receive treatment (which may be the prevailing policy when an available treatment is believed to be uniformly effective) and to a rule that recommends no one receive treatment (which might be the preferred strategy when the effectiveness of a proposed treatment has not yet been established).

Table \ref{tab:simulation_results_mean_response} shows the advantage of using split-regression with IPW weights relative to the ``naive'' uniform weighting: 
The bias of the naive approach forces the estimation of the incorrect non-crossing response-curves shown with the solid lines in the left panel of Figure \ref{fig:simulations_one_scenario}. In contrast, split-regression with the correct IPW weighting approaches the optimal treatment rule with large enough sample size because it is estimating the crossing response-curves in the right panel of Figure \ref{fig:simulations_one_scenario}. The code used to perform the simulations is shared at \texttt{github.com/jhroth/simulations-split-regression}.

\section{Data Example: WHI-OS} \label{obs_sec:data_example}
We also illustrate the split-regression approach and compare it to alternatives by applying the \texttt{R} package \texttt{DevTreatRules} to the Women's Health Initiative Observational Study component (WHI-OS). A detailed description of the study design and summaries of baseline measurements for participants (postmenopausal women between the ages of 49 and 81 who were recruited for the WHI clinical trial but either declined to participate or were later deemed ineligible) are available in \cite{Langer:2003}. We also present summary tables of the variables we retained for analysis in the appendix. The GitHub page \texttt{github.com/jhroth/data-examplesplit-regression} contains the \texttt{R} code needed to go start-to-finish from loading the raw WHI-OS datasets (access to which requires additional permission) to replicating our estimates in the evaluation subset.

Briefly, we aim to build a treatment rule to assign baseline hormone therapy (HRT) --  defined 
as \emph{currently using} unopposed estrogen and/or estrogen plus progesterone at baseline -- to postmenopausal women if it will increase a woman's probability of remaining free of coronary heart disease (CHD) after 10 years and, in a separate analysis, if it will increase a woman's probability of remaining free of breast cancer after 10 years. All variables included in our analysis besides the outcomes were measured at baseline. We used the ``adjudicated'' outcome variables in the WHI-OS as described in \cite{Curb:2003}.

We classified $4$ categorical variables as belonging to \CTN: education level, ethnicity, family income, and how each participant heard about the study. We identified $31$ self-reported variables to make up \CTI and we chose $\RR=\CTI$, so there are $31$ candidates to be used as inputs in our treatment rule. We did not specify any variables as belonging to \CNI. In the Appendix, we present the complete list of these variables and simple summaries including counts of missing values. We intend for this to be primarily an illustrative example rather than a definitive claim about appropriate variable classifications in this study.

About 17.6\% of the $94140$ observations in the initial dataset had a missing value of at least one variable in $\CT$ or \RR.  Instead of conducting a complete-case analysis that would drop observations with missing values, we used an IPW-based adjustment for missingness \citep{Seaman:2013}. Our shared code 
shows the details of our adjustment for missingness using the \texttt{additional.weights} argument of the \texttt{BuildRule()} and \texttt{EvaluateRule()} functions in \texttt{DevTreatRules}.

The top-half of Table \ref{tab:selected_rules_validation_and_evaluation} presents estimated ATEs and ABR in the validation set for two split-regression specifications using the outcome of no breast cancer after 10 years: one that used ridge regression for the propensity score and lasso for the rule, and another specification that used logistic regression for both the propensity score and rule models. On the validation set, none of the split-regression specifications for the outcome of no CHD after 10 years appeared to be an improvement over the naive strategy of treating no one and thus none of those specifications were chosen. Since the estimated ATEs and ABR in Table \ref{tab:selected_rules_validation_and_evaluation} informed our model selection -- in particular, we chose these models because their ABRs in the validation set were relatively high -- they do not serve as trustworthy estimates of ATE and ABR in independent samples drawn from this population in the future.

The bottom-half of Table \ref{tab:selected_rules_validation_and_evaluation} presents the estimated ATEs and ABR in the evaluation set, which did not inform model selection. 
Unfortunately, the 95\% CI for estimated ATE among the treated population contains $0$ for both rules in the evaluation set; as a result, we do not find evidence that either rule has identified a subpopulation of individuals who appear to benefit from treatment.

\section{Discussion}\label{obs_sec:discussion}
We outlined a principled approach to classify the roles of variables collected in an observational study into clinically meaningful categories and, using that knowledge, to develop a treatment rule along with a trustworthy estimate of the rule's population benefit. Since this paper is intended to be a practical guide to help practitioners go from start to finish in estimating and evaluating treatment rules without getting bogged down by the onerous and error-prone tasks of coding the method from scratch in statistical software, we implemented our approach in the R package \texttt{DevTreatRules} and shared the code needed to reproduce our simulations and data example on GitHub.

In a simple simulation study, we saw the benefit of estimating a treatment rule using this paper's preferred approach with IPW weighting compared to using uniform observational weights that ignore the observational study design. In the WHI-OS data example, we used the split-regression approach to develop a treatment rule that assigns baseline hormone therapy to postmenopausal women if it is expected to increase their probability of remaining free of coronary heart disease after 10 years or free of breast cancer after 10 years. With the 10-year CHD outcome, split-regression did not estimate a rule with a positive estimate of ATE in the treated subgroup on the validation set. With the 10-year breast cancer outcome, however, split-regression did identify a rule that recommends HRT to about 18\% of women and, among this treated subpopulation, had an estimated 1.5 percentage-point decrease in the probability of breast cancer. However, this 1.5 percentage-point decrease lacks statistical significance (95\% CI: $-0.027, 0.068$) and, as a result, we would simply recommend not assigning HRT to any women in this population if the goal is to reduce 10-year breast cancer incidence or 10-year CHD incidence. This null finding is, unfortunately, often a typical result in the search for informative treatment rules with observational data.

One notable limitation of this work is that, while the algorithm outlined in Section \ref{obs_subsec:recipe} is fairly general, the accompanying \texttt{R} implementation does not have as much flexibility (e.g. the currently available estimation methods are linear/logistic regression and its lasso/ridge counterparts); in future work we hope to expand the package to support more estimation methods. Another limitation is that our data example is not publicly reproducible because we are unable to share the underlying WHI-OS dataset; however, we do share the code that will reproduce the data example for users who have access to the raw WHI-OS data files.

\singlespacing

\bibliographystyle{apalike}
\bibliography{uwthesis}

\newpage

\newpage
\begin{table}[H]
  \begin{center}
    \begin{tabular}{ccc} 
      \hline
      \textbf{Potentially influences treatment} & \multicolumn{2}{c}{\textbf{Observed in}} \\
      \textbf{in current study} & \multicolumn{2}{c}{\textbf{independent clinical settings}} \\
      & Yes & No \\
      \hline
      Yes & \CTI  & \CTN \\
      No & \CNI & \CNN \\ 
      \hline
    \end{tabular}
    \caption{Proposed partitioning of individual characteristics in an observational study}
    \label{tab:variable_roles}
  \end{center}
\end{table}

\begin{table}[H]
  \resizebox{\columnwidth}{!}{%
  \begin{tabular}{l p{2cm} p{2.5cm} p{2.5cm} p{2.5cm} p{1.5cm} p{2.5cm}}
    \hline
    \multirow{3}{*}[-2.5em]{\textbf{Framework}} & 
    Type of approach & Accommodates observational data & Distinguishes between $(\CTI, \CNI)$ and $(\CTN)$ & Accommodates a range of statistical learning methods & R package & Shares code to reproduce simulations or application \\
    \hline 
    \cite{Zhao:2012} & Direct & \textcolor{blue}{Yes}  & \textcolor{orange}{No} & \textcolor{orange}{No} & \textcolor{blue}{Yes}$^1$ & \textcolor{orange}{No} \\
    \cite{Zhang:2012, Zhang_robust:2012} & Direct & \textcolor{blue}{Yes} & \textcolor{orange}{No} & \textcolor{blue}{Yes} & \textcolor{orange}{No} & \textcolor{blue}{Yes} \\
    \cite{Zhang:2015} & Direct & \textcolor{blue}{Yes} & \textcolor{orange}{No} & \textcolor{orange}{No} & \textcolor{orange}{No} & \textcolor{blue}{Yes} \\
    \cite{Qian:2011} & Direct & \textcolor{orange}{No} & \textcolor{orange}{No} & \textcolor{orange}{No} & \textcolor{orange}{No} & \textcolor{orange}{No} \\
    \cite{Chen:2017} & Direct & \textcolor{blue}{Yes} & \textcolor{orange}{No} & \textcolor{blue}{Yes} & \textcolor{blue}{Yes}$^2$ & \textcolor{blue}{Yes} \\
    \cite{Cai:2011} & Indirect & \textcolor{orange}{No} & \textcolor{orange}{No} & \textcolor{blue}{Yes} & \textcolor{orange}{No} & \textcolor{orange}{No} \\
    \cite{Lu:2013} & Indirect & \textcolor{orange}{No} & \textcolor{orange}{No} & \textcolor{blue}{Yes} & \textcolor{orange}{No} & \textcolor{orange}{No} \\ 
    \cite{Kang:2014} & Indirect & \textcolor{orange}{No} & \textcolor{orange}{No} & \textcolor{orange}{Yes} & \textcolor{orange}{No} & \textcolor{blue}{Yes} \\
    \cite{McKeague:2014} & Indirect & \textcolor{orange}{No} & \textcolor{orange}{No} & \textcolor{blue}{Yes} & \textcolor{orange}{No} & \textcolor{orange}{No} \\
    \cite{Ciarleglio:2015} & Indirect & \textcolor{blue}{Yes} & \textcolor{orange}{No} & \textcolor{orange}{No} & \textcolor{orange}{No} & \textcolor{blue}{Yes} \\ 
    &&&&&& \\
    \textbf{\emph{This paper}} & \textbf{\emph{Indirect}} & \emph{\textbf{\textcolor{blue}{Yes}}} & \emph{\textbf{\textcolor{blue}{Yes}}} & \emph{\textbf{\textcolor{blue}{Yes}}} & \emph{\textbf{\textcolor{blue}{Yes}}} & \emph{\textbf{\textcolor{blue}{Yes}}} \\
    \hline 
  \end{tabular}
  }
  \caption{Selected methods for estimating treatment rule \\
              $^1$\cite{DynTxRegime} \\
              $^2$\cite{personalized}}
  \label{tab:selected_methods}
\end{table}

\begin{table}[H]
  \begin{center}
  \begin{tabular}{lrrrrr}
    \hline
    & \multicolumn{5}{c}{\textbf{Sample size in development set}} \\
   \textbf{Type of Rule} & 50 & 100 & 200 & 500 & 1000 \\
    \hline
    Split-regression & 0.543 & 0.553
    & 0.562 & 0.57 
    & 0.572 \\
    Split-regression (naive, no weights) & 0.552 & 0.554
    & 0.553 & 0.55 
    & 0.549 \\
    \hline
    \hline
    Optimal rule & \textbf{0.574} & \textbf{0.574}
    & \textbf{0.574} & \textbf{0.574} 
    & \textbf{0.574} \\
    Treating all & 0.479 & 0.479
    & 0.479 & 0.479 
    & 0.479 \\
    Treating none & 0.543 & 0.543
    & 0.543 & 0.543 
    & 0.543 \\
    \hline
   \end{tabular}
  \caption{Mean outcome probability, as a function of rule type and development set sample size, averaged over 1000 replications and calculated in evaluation sets of size 10000}
  \label{tab:simulation_results_mean_response}
 \end{center}
\end{table}

\begin{table}[H]
  \begin{center}
    \resizebox{\columnwidth}{!}{%
   \begin{tabular}{lrrrrr}
     \hline
     & \textbf{Positives} & \textbf{Negatives} & \textbf{ATE in Positives} & \textbf{ATE in Negatives} & \textbf{ABR} \\
     \hline 
     \textbf{Estimates on Validation Set} \\
     \hspace{3mm} Split regression (ridge/lasso) & 
     3544 &
     15758 &
     0.013 &
     -0.051 &
     0.044 \\
     \hspace{3mm} Split regression (logistic/logistic) & 
     3692 &
     15610 &
     0.021 &
     -0.053 &
     0.047 \\
     \hspace{3mm} Treat no one (logistic/NA) & 
     0 &
     19302 &
     NA &
     -0.045 &
     0.045 \\ 
     \hline
     \textbf{Estimates on Evaluation Set} \\
     \hspace{3mm} Split-regression (ridge/lasso) &
     3408 &
     15894 &
     0.015 (-0.027, 0.068) & 
     -0.045 (-0.067, -0.031) &
     0.04 \\
     \hspace{3mm} Split-regression (logistic/logistic) &
     3569 &
     15733 &
     0.002 (-0.049, 0.052) & 
     -0.048 (-0.07, -0.035) & 
     0.04 \\
     \hspace{3mm} Treat no one (logistic/NA) & 
     0 &
     19302 &
     NA & 
     -0.05&
     0.05 \\
     \hline
  \end{tabular}
  }
    \caption{Summary of Selected Rules on the Validation and Evaluation  Sets (Outcome: No breast cancer after 10 years). The selected propensity method/rule method are in parentheses}
  \label{tab:selected_rules_validation_and_evaluation}
  \end{center}
\end{table}

\newpage
\appendix
\begin{center}
\Large{Appendix: \\ Using Propensity Scores to Develop and Evaluate Treatment Rules with Observational Data}
\end{center}
\section{Motivation for IPW Estimator of Average Treatment Effect} \label{appendix_sec:IP}
We are starting with
\begin{equation}
  \Uppsi_\RR \equiv \int_{\CT} \left \{\textcolor{blue}{E(Y \mid \CT, \RR, T=1)} - \textcolor{red}{E(Y \mid \CT, \RR, T=0)} \right \} \: dP^{\theo}(\CT \mid \RR), \label{eq:define_ATE_regression}
\end{equation}
and, focusing on the $T=1$ group for a fixed $\RR$, we can write
{\footnotesize
\begin{align}
  \Uppsi (T=1 \mid \RR) = &\int_{\CT} \textcolor{blue}{E(Y \mid \CT, \RR, T=1)} \: dP^{\theo}(\CT \mid \RR), \\
  &= \int_{\CT} \textcolor{blue}{\min_{f \in \FF} \left\{ \int_{\YY} \left(y - f(\RR)\right)^2 \: dP^{\act}(Y \mid \CT, \RR, T=1)\right\}} \: dP^{\theo}(\CT \mid \RR), \\
  &= \min_{f \in \FF} \int_{\CT} \int_{\YY} \left(y - f(\RR)\right)^2 \: dP^{\act}(Y \mid \CT, \RR, T=1) \: dP^{\theo}(\CT \mid \RR).
\end{align}
}
For random $\RR$, we have
{\scriptsize
\begin{align}
  \Uppsi (T=1, \RR) \equiv & \int_{\RR} \int_{\CT}\textcolor{blue}{E(Y \mid \CT, \RR, T=1)} \: dP^{\theo}(\CT \mid \RR) \:dP^{\act}(\RR \mid T=1) \nonumber \\
&= \int_{\RR} \int_{\CT} \textcolor{blue}{\min_{f \in \FF}\bigg\{ \int_{\YY}\left(y - f(\RR) \right)^2 \: dP^{\act}(Y \mid \CT, \RR, T=1) \bigg\}} \: dP^{\theo}(\CT \mid \RR) \: dP^{\act}(\RR \mid T=1) \nonumber \\
&= \int_{\RR} \min_{f \in \FF} \int_{\CT} \int_{\YY} \left(y - f(\RR)\right)^2 \: dP^{\theo}(\CT \mid \RR) \: d P^{\act}(Y \mid \CT, \RR, T=1) \: dP^{\act}(\RR \mid T=1) \nonumber \\
&= \int_{\RR} \min_{f \in \FF} \int_{\CT} \int_{\YY} \left(y - f(\RR)\right)^2 \: dP^{\theo}(\CT \mid \RR) \: d P^{\act}(Y \mid \CT, \RR, T=1) \: dP^{\act}(\RR \mid T=1) \: \frac{dP^{\act}(\CT \mid \RR, T=1)}{dP^{\act}(\CT \mid \RR, T=1)} \nonumber \\
&= \int_{\RR} \min_{f \in \FF} \int_{\CT} \int_{\YY} \left(y - f(\RR)\right)^2 \: \frac{dP^{\theo}(\CT \mid \RR)}{dP^{\act}(\CT \mid \RR, T=1)} \: d P^{\act}(Y, \CT, \RR \mid T=1) \label{eq:regression_before_bayes}.
\end{align}
}
Since
\begin{align*}
  P^{\act}(\CT \mid \RR, T=1) &= \frac{P^{\act}(\CT, \RR, T=1)}{P^{\act}(T=1 \mid \RR) P^{\act}(\RR)} \\
&= \frac{P^{\act}(T=1 \mid \CT, \RR) P^{\act}(\CT \mid \RR) P^{\act}(\RR)}{P^{\act}(T=1 \mid \RR) P^{\act}(\RR)}, \\
&= \frac{P^{\act}(T=1 \mid \CT, \RR) P^{\act}(\CT \mid \RR)}{P^{\act}(T=1 \mid \RR)}, \\
&\equiv \frac{\pi(\CT, \RR) P^{\act}(\CT \mid \RR)}{\pi(\RR)},
\end{align*}
we can rewrite (\ref{eq:regression_before_bayes}) as

\begin{align*}
\Uppsi (T=1, \RR) &\equiv \int_{\RR} \min_{f \in \FF} \int_{\CT} \int_{\YY} \left(y - f(\RR)\right)^2 \: \frac{dP^{\theo}(\CT \mid \RR)}{dP^{\act}(\CT \mid \RR, T=1)} \: d P^{\act}(Y, \CT, \RR \mid T=1) \\
&= \int_{\RR} \min_{f \in \FF} \int_{\CT} \int_{\YY} \left(y - f(\RR)\right)^2 \frac{dP^{\theo}(\CT \mid \RR) \pi(\RR)}{\pi(\CT, \RR) dP^{\act}(\CT \mid \RR)} \: dP^{\act}(Y, \CT, \RR \mid T=1) \\
&= \int_{\RR} \min_{f \in \FF} \int_{\CT} \int_{\YY} \left(y - f(\RR)\right)^2 \frac{dP^{\theo}(\CT \mid \RR)}{dP^{\act}(\CT \mid \RR)} \cdot \frac{\pi(\RR)}{\pi(\CT, \RR)} \: dP^{\act}(Y, \CT, \RR \mid T=1) 
\end{align*}

If we assume $dP^{\theo}(\CT \mid \RR) = dP^{\act}(\CT \mid \RR)$ (i.e. representative sampling of covariates, conditional on values of the biomarkers), then this becomes
\begin{align}
\Uppsi (T=1, \RR) = \int_{\RR} \min_{f \in \FF} \int_{\CT} \int_{\YY} \left(y - f(\RR)\right)^2 \frac{\pi(\RR)}{\pi(\CT, \RR) } \: dP^{\act}(Y, \CT, \RR \mid T=1) \label{eq:regression_before_PIE}.
\end{align}
If we assume the propensity scores are known, then the plug-in estimator of (\ref{eq:regression_before_PIE}) is
\begin{align}
\min_{f \in \FF} \frac{1}{N_1} \sum_{i=1}^N I(T_i = 1) \: \frac{\pi(\RR_i)}{\pi(\CT_i, \RR_i)} \: \left[Y_i - f(\RR_i)\right]^2  \label{eq:regression_PIE},
\end{align}
where $N_1 = \sum_{i=1}^N I(T_i = 1)$. We note that (\ref{eq:regression_PIE}) is weighted squared-error loss among $T=1$ group with weights $w_1(\CT, \RR) \equiv \pi(\RR) / \pi(\CT, \RR)$. \\ \\
Similarly in the $T=0$ group, the target parameter is
{\footnotesize
\begin{align*}
  \Uppsi (T=0, \RR) &\equiv \int_{\RR} \textcolor{red}{E(Y \mid \CT, \RR, T=0)} \: \: dP^{\theo}(\CT \mid \RR) \:dP^{\act}(\RR \mid T=1) \\
&= \int_{\RR} \int_{\CT} \textcolor{red}{\min_{f \in \FF}\bigg\{ \int_{\YY}\left(y - f(\RR) \right)^2 \: dP^{\act}(Y \mid \CT, \RR, T=0) \bigg\}} \: dP^{\theo}(\CT \mid \RR) \: dP^{\act}(\RR \mid T=0) \\
&= \int_{\RR} \min_{f \in \FF} \int_{\CT} \int_{\YY} \left(y - f(\RR)\right)^2 \frac{1 - \pi(\RR)}{1 - \pi(\CT, \RR) } \: dP^{\act}(Y, \CT, \RR \mid T=0),
\end{align*}
}
whose plug-in estimator, assuming the propensity scores are known, is
\begin{align}
\min_{f \in \FF} \frac{1}{N_0} \sum_{i=1}^N I(T_i = 0) \: \frac{1 - \pi(\RR_i)}{1- \pi(\CT_i, \RR_i)} \: \left[Y_i - f(\RR_i)\right]^2 \label{eq:regression_PIE_controls} 
\end{align}
where $N_0 = \sum_{i=1}^N I(T_i = 0)$. We note that (\ref{eq:regression_PIE}) is weighted squared-error loss among $T=0$ group with weights $w_0(\CT, \RR) \equiv (1 - \pi(\RR)) / (1 - \pi(\CT, \RR))$.

\section{Evaluating the Rule}
As described in the paper, a foundational target parameter in our framework is the average treatment effect (ATE) in a subpopulation of individuals with characteristics $\RR=\rr$.
\begin{equation}
  E[Y^{1} - Y^{0} \mid \RR=\rr], \label{eq:conditioned_ATE_again}
\end{equation}
which under the assumptions of consistency, no unmeasured confounding, and positivity, as detailed in \cite{Kennedy:2015} can be re-written as
\begin{equation}
  \resizebox{.90 \textwidth}{!} 
  {
    $\psi(\rr) \equiv \int_\CT \left\{E\left[Y \mid \CT, \RR=\rr,T=1\right] - E\left[Y \mid \CT, \RR=\rr, T=0 \right]\right\}dP(\CT \mid \RR=\rr)$ \label{eq:target_ATE_evaluating},
    }
\end{equation}
which implies we should recommend treatment to individuals in the subgroup
\begin{equation}
  \Omega^+ = \left\{\rr\mid \psi(\rr) > 0 \right\} \label{eq:population_to_treat_evaluating},
\end{equation}
which is known as the \emph{test-positives} group. If $\RR$ were a set of gene expression levels, for example, then $\Omega^+$ would be the subset of gene expression levels for which treatment increases the expected number of months until relapse. Again, the expected improvement in months until relapse among this treated subpopulation is
\begin{equation}
\int_{\rr\in\Omega^+} \psi(\rr) dP(\rr) \label{eq:population_treatment_benefit}.
\end{equation}
Similarly, treatment should not be recommended to individuals in the \emph{test-negatives} subpopulation defined by
\begin{equation}
  \Omega^- = \left\{\rr\mid \psi(\rr) \leq 0 \right\} \label{eq:population_to_avoid_treatment},
\end{equation}
and
\begin{equation}
\int_{\rr\in\Omega^-} \psi(\rr) dP(\rr) \label{eq:population_avoid_treatment_benefit}.
\end{equation}
would yield the average increase in months until relapse for the subpopulation that avoids treatment.

To obtain a trustworthy estimate of how a developed treatment rule will perform for individuals seen in future clinical settings, it is absolutely essential that development of the rule and evaluation of the selected rule are performed independently. Chapter 7 (Model Assessment and Selection) of \cite{Hastie:2008} provides an excellent discussion of this topic with regard to the predictive performance of model-based estimates. In brief, if the development and evaluation datasets coincide then our evaluation of the treatment rule's benefit will be overly optimistic because it will reward the rule for incorrectly classifying noise in the development dataset as signal that would persist when we apply the rule to future individuals.

We can take the treatment rule $\tilde{B}(\rr)$ -- which gives us a mapping from an individual's particular characteristics $\RR=\rr$ to a treatment recommendation -- that was estimated on a \emph{development} dataset and apply it to the independent \emph{evaluation dataset} to yield, for the $n_{\text{eval}}$ individuals in the evaluation dataset indexed by $i=1, \ldots, n_{\text{eval}}$,
\begin{equation}
  \tilde{\Omega}^+ = \left\{\rr_i\mid \tilde{B}(\rr_i) = 1 \right\}, \label{eq:estimator_population_to_treat}
\end{equation}
the test-positives subset of observations in the evaluation dataset, based on $\RR$, who are expected to have more months until relapse under treatment than under standard-of-care. Similarly,
\begin{equation}
  \tilde{\Omega}^- = \left\{\rr_i\mid \tilde{B}(\rr_i) = 0 \right\} \label{eq:estimator_population_to_not_treat}
\end{equation}
yields the test-negatives subset of observations in the evaluation dataset, based on $\RR$, who are expected to have fewer months until relapse under treatment than under standard-of-care.

\section{Data Example: Summary of Dataset} \label{appendix_sec:data_example_summary}

\begin{table}[H]
  \caption{Summary of Outcome and Treatment Variables}
  \vspace{-5mm}
  \begin{center}
    \resizebox{\columnwidth}{!}{%
  \begin{tabular}{lllll}
    \hline
    & \multicolumn{1}{c}{\textbf{Overall (93676)}} & \multicolumn{1}{c}{\textbf{Non-event (3063)}} & \multicolumn{1}{c}{\textbf{Event (90613)}} & \textbf{N missing}\\
    \hline 
    \textbf{Outcomes} &&&& \\
     \hspace{5mm} No CHD within 10 years of enrollment, n (\%) & 90613 (97\%) &
     - & - & 0 \\
     \hspace{5mm} No breast cancer within 10 years of enrollment, n (\%) & 88883 (95\%) &
     - & - & 0 \\
    \textbf{Treatment} &&&& \\
    \hspace{5mm} Currently using unopposed estrogen and/or estrogen plus progesterone, n (\%) & 41630 (44\%) &
    1003 (33\%)  &
    40627 (45\%) & 85 \\
    \hline
  \end{tabular}
    }
  \end{center}
\end{table}        

\begin{table}[H]
  \vspace{-5mm}
  \begin{center}
    \resizebox{\columnwidth}{!}{%
  \begin{tabular}{lllll}
    \hline
    & \multicolumn{1}{c}{\textbf{Overall (93676)}} & \multicolumn{1}{c}{\textbf{Non-event (3063)}} & \multicolumn{1}{c}{\textbf{Event (90613)}} & \textbf{N missing}\\
    \hline 
    \textbf{Highest grade completed} &&&& \textbf{767}\\
    \hspace{5mm} None, n (\%) & 84 (0\%) &
    2 (0\%)  &
    82 (0\%) & \\
    \hspace{5mm} 1-4, n (\%) & 356 (0\%) &
    11 (0\%)  &
    345 (0\%) & \\
    \hspace{5mm} 5-8, n (\%) & 1121 (1\%) &
    51 (2\%)  &
    1070 (1\%) & \\
    \hspace{5mm} 9-11, n (\%) & 3288 (4\%) &
    184 (6\%)  &
    3104 (3\%) & \\
    \hspace{5mm} High school, n (\%) & 15122 (16\%) &
    592 (19\%)  &
    14530 (16\%) & \\
    \hspace{5mm} Vocational, n (\%) & 9123 (10\%) &
    369 (12\%)  &
    8754 (10\%) & \\
    \hspace{5mm} Some college, n (\%) & 24812 (27\%) &
    828 (27\%)  &
    23984 (27\%) & \\
    \hspace{5mm} College, n (\%) & 10669 (11\%) &
    277 (9\%)  &
    10392 (12\%) & \\
    \hspace{5mm} Some post-graduate, n (\%) & 11018 (12\%) &
    314 (10\%)  &
    10704 (12\%) & \\
    \hspace{5mm} Master's, n (\%) & 14732 (16\%) &
    343 (11\%)  &
    14389 (16\%) & \\
    \hspace{5mm} Doctoral, n (\%) & 2584 (3\%) &
    67 (2\%)  &
    2517 (3\%) & \\
    \textbf{Ethnicity} &&&& \textbf{265}\\
    \hspace{5mm} American Indian or Alaskan Native, n (\%) & 421 (0\%) &
    18 (1\%)  &
    403 (0\%) & \\
    \hspace{5mm} Asian or Pacific Islander, n (\%) & 2671 (3\%) &
    52 (2\%)  &
    2619 (3\%) & \\
    \hspace{5mm} Black or African-American, n (\%) & 7635 (8\%) &
    283 (9\%)  &
    7352 (8\%) & \\
    \hspace{5mm} Hispanic/Latino, n (\%) & 3609 (4\%) &
    56 (2\%)  &
    3553 (4\%) & \\
    \hspace{5mm} White (non-Hispanic), n (\%) & 78016 (84\%) &
    2611 (86\%)  &
    75405 (83\%) & \\
    \hspace{5mm} Other, n (\%) & 0 (0\%) &
    28 (1\%)  &
    1031 (1\%) & \\
    \textbf{Heard about study} &&&& \textbf{1281}\\
    \hspace{5mm} Mailed letter, n (\%) & 47623 (52\%) &
    1722 (57\%)  &
    45901 (51\%) & \\
    \hspace{5mm} Brochure, n (\%) & 9789 (11\%) &
    317 (10\%)  &
    9472 (11\%) & \\
    \hspace{5mm} TV, n (\%) & 2731 (3\%) &
    93 (3\%)  &
    2638 (3\%) & \\
    \hspace{5mm} Radio, n (\%) & 1017 (1\%) &
    24 (1\%)  &
    993 (1\%) & \\
    \hspace{5mm} Newspaper or magaize, n (\%) & 14610 (16\%) &
    415 (14\%)  &
    14195 (16\%) & \\
    \hspace{5mm} Meeting, n (\%) & 1158 (1\%) &
    32 (1\%)  &
    1126 (1\%) & \\
    \hspace{5mm} Friend or relative, n (\%) & 9408 (10\%) &
    223 (7\%)  &
    9185 (10\%) & \\
    \hspace{5mm} Other, n (\%) & 6059 (7\%) &
    198 (7\%)  &
    5861 (7\%) & \\
    \textbf{Family income} &&&& \textbf{4119}\\
    \hspace{5mm} Less than \$10,000, n (\%) & 3917 (4\%) &
    248 (8\%)  &
    3669 (4\%) & \\
    \hspace{5mm} \$10,000 - \$19,999, n (\%) & 10101 (11\%) &
    504 (17\%)  &
    9597 (11\%) & \\
    \hspace{5mm} \$20,000 - \$34,999, n (\%) & 20226 (23\%) &
    838 (29\%)  &
    19388 (22\%) & \\
    \hspace{5mm} \$35,000 - \$49,999, n (\%) & 17430 (19\%) &
    536 (18\%)  &
    16894 (19\%) & \\
    \hspace{5mm} \$50,000 - \$74,999, n (\%) & 17487 (20\%) &
    409 (14\%)  &
    17078 (20\%) & \\
    \hspace{5mm} \$75,000 - \$99,999, n (\%) & 8181 (9\%) &
    169 (6\%)  &
    8012 (9\%) & \\
    \hspace{5mm} \$100,000 - \$149,999, n (\%) & 6034 (7\%) &
    73 (3\%)  &
    5961 (7\%) & \\
    \hspace{5mm} \$150,000 or more, n (\%) & 3393 (4\%) &
    42 (1\%)  &
    3351 (4\%) & \\
    \hspace{5mm} Don't know, n (\%) & 2788 (3\%) &
    99 (3\%)  &
    2689 (3\%) & \\
    \hline
  \end{tabular}
  }
  \end{center}
  \caption{Summary of Variables Influencing Only Treatment Assignment}
\end{table}    

\begin{table}[H]
  \vspace{-5mm}
  \begin{center}
    \scriptsize{
  \begin{tabular}{lllll}
    \hline
    & \multicolumn{1}{c}{\textbf{Overall (93676)}} & \multicolumn{1}{c}{\textbf{Non-event (3063)}} & \multicolumn{1}{c}{\textbf{Event (90613)}} & \textbf{N missing}\\
    \hline 
    Age, mean (IQR) & 63.6 (58, 69) &
    68.3 (64, 73) &
    63.5 (57, 69) & 0 \\
    Angina ever, n (\%) & 5547 (6\%) &
    584 (19\%)  &
    4963 (6\%) & 708 \\
    Aortic aneurysm ever, n (\%) & 187 (0\%) &
    30 (1\%)  &
    157 (0\%) & 1523 \\
    Breast cancer ever, n (\%) & 5299 (6\%) &
    208 (7\%)  &
    5091 (6\%) & 879 \\
    Coronary bypass surgery ever, n (\%) & 881 (1\%) &
    204 (7\%)  &
    677 (1\%) & 1513 \\
    Cancer ever, n (\%) & 12075 (13\%) &
    481 (16\%)  &
    11594 (13\%) & 752 \\
    Cardiac catheterization ever, n (\%) & 3837 (4\%) &
    453 (15\%)  &
    3384 (4\%) & 1513 \\
    Carotid endarterectomy/angioplasty ever, n (\%) & 344 (0\%) &
    59 (2\%)  &
    285 (0\%) & 1510 \\
    Cervix cancer ever, n (\%) & 1205 (1\%) &
    44 (1\%)  &
    1161 (1\%) & 916 \\
    Heart failure ever, n (\%) & 893 (1\%) &
    134 (4\%)  &
    759 (1\%) & 7 \\
    Cadiovascular disease ever, n (\%) & 17523 (19\%) &
    1206 (40\%)  &
    16317 (18\%) & 2045 \\
    Hysterectomy ever, n (\%) & 39149 (42\%) &
    1415 (46\%)  &
    37734 (42\%) & 87 \\
    Diabetes ever, n (\%) & 5318 (6\%) &
    544 (18\%)  &
    4774 (5\%) & 96 \\
    Stroke ever, n (\%) & 1415 (2\%) &
    142 (5\%)  &
    1273 (1\%) & 56 \\
    \textbf{Have a lot of energy?} &&&& \textbf{772}\\
    \hspace{5mm} All the time, n (\%) & 4740 (5\%) &
    108 (4\%)  &
    4632 (5\%) & \\
    \hspace{5mm} Most the time, n (\%) & 33633 (36\%) &
    766 (25\%)  &
    32867 (37\%) & \\
    \hspace{5mm} A good bit, n (\%) & 20668 (22\%) &
    642 (21\%)  &
    20026 (22\%) & \\
    \hspace{5mm} Some times, n (\%) & 19397 (21\%) &
    780 (26\%)  &
    18617 (21\%) & \\
    \hspace{5mm} A little bit, n (\%) & 9956 (11\%) &
    481 (16\%)  &
    9475 (11\%) & \\
    \hspace{5mm} Never, n (\%) & 4510 (5\%) &
    258 (9\%)  &
    4252 (5\%) & \\
    \textbf{General health} &&&& \textbf{655}\\
    \hspace{5mm} Excellent, n (\%) & 16576 (18\%) &
    263 (9\%)  &
    16313 (18\%) & \\
    \hspace{5mm} Very good, n (\%) & 37684 (41\%) &
    861 (28\%)  &
    36823 (41\%) & \\
    \hspace{5mm} Good, n (\%) & 29669 (32\%) &
    1280 (42\%)  &
    28389 (32\%) & \\
    \hspace{5mm} Fair, n (\%) & 8210 (9\%) &
    550 (18\%)  &
    7660 (9\%) & \\
    \hspace{5mm} Poor, n (\%) & 882 (1\%) &
    82 (3\%)  &
    800 (1\%) & \\
    High cholesterol requiring pills ever, n (\%) & 13773 (15\%) &
    748 (25\%)  &
    13025 (15\%) & 2071 \\
    \textbf{Hot flash in past 4 weeks} &&&& \textbf{733}\\
    \hspace{5mm} No, n (\%) & 15158 (16\%) &
    386 (13\%)  &
    14772 (16\%) & \\
    \hspace{5mm} Mild, n (\%) & 4593 (5\%) &
    139 (5\%)  &
    4454 (5\%) & \\
    \hspace{5mm} Moderate, n (\%) & 1267 (1\%) &
    38 (1\%)  &
    1229 (1\%) & \\
    \hspace{5mm} Severe, n (\%) & 0 (0\%) &
    0 (0\%)  &
    0 (0\%) & \\
    \textbf{Hypertension} &&&& \textbf{1699}\\
    \hspace{5mm} Never, n (\%) & 61196 (67\%) &
    1270 (42\%)  &
    59926 (67\%) & \\
    \hspace{5mm} Yes, untreated, n (\%) & 7317 (8\%) &
    338 (11\%)  &
    6979 (8\%) & \\
    \hspace{5mm} Yes, treated, n (\%) & 23464 (26\%) &
    1392 (46\%)  &
    22072 (25\%) & \\
    \textbf{Recent physical/emotional problems socially} &&&& \textbf{747}\\
    \hspace{5mm} Not at all, n (\%) & 68565 (74\%) &
    2004 (66\%)  &
    66561 (74\%) & \\
    \hspace{5mm} Slightly, n (\%) & 14249 (15\%) &
    549 (18\%)  &
    13700 (15\%) & \\
    \hspace{5mm} Moderately, n (\%) & 6121 (7\%) &
    279 (9\%)  &
    5842 (6\%) & \\
    \hspace{5mm} Quite a bit, n (\%) & 3217 (3\%) &
    169 (6\%)  &
    3048 (3\%) & \\
    \hspace{5mm} Extremely, n (\%) & 777 (1\%) &
    34 (1\%)  &
    743 (1\%) & \\
    Quality of life (1-10), mean (IQR) & 8.3 (8, 9) &
    8.1 (7, 9) &
    8.3 (8, 9) & 724 \\
    Menopause before age 40, n (\%) & 8352 (9\%) &
    339 (12\%)  &
    8013 (9\%) & 3951 \\
    MENPSYMP, n (\%) & 64608 (71\%) &
    1922 (65\%)  &
    62686 (71\%) & 2425 \\
    \textbf{Limited in daily activities?} &&&& \textbf{726}\\
    \hspace{5mm} Yes, limited a lot, n (\%) & 6263 (7\%) &
    442 (15\%)  &
    5821 (6\%) & \\
    \hspace{5mm} Yes, limited a little, n (\%) & 23110 (25\%) &
    1132 (37\%)  &
    21978 (24\%) & \\
    \hspace{5mm} No, not limited at all, n (\%) & 63577 (68\%) &
    1459 (48\%)  &
    62118 (69\%) & \\
    \textbf{One or both ovaries removed} &&&& \textbf{552}\\
    \hspace{5mm} No, n (\%) & 65240 (70\%) &

    2019 (66\%)  &
    63221 (70\%) & \\
    \hspace{5mm} Yes, one taken out, n (\%) & 6583 (7\%) &
    217 (7\%)  &
    6366 (7\%) & \\
    \hspace{5mm} Yes, both taken out, n (\%) & 18890 (20\%) &
    713 (23\%)  &
    18177 (20\%) & \\
    \hspace{5mm} Yes, unknown number taken out, n (\%) & 738 (1\%) &
    29 (1\%)  &
    709 (1\%) & \\
    \hspace{5mm} Yes, part of ovary taken out, n (\%) & 893 (1\%) &
    30 (1\%)  &
    863 (1\%) & \\
    \hspace{5mm} Don't know, n (\%) & 780 (1\%) &
    36 (1\%)  &
    744 (1\%) & \\
    Osteoporosis ever, n (\%) & 8282 (9\%) &
    385 (13\%)  &
    7897 (9\%) & 1240 \\
    Any part of ovaries removed before age 40, n (\%) & 8279 (9\%) &
    301 (10\%)  &
    7978 (9\%) & 1120 \\
    Peripheral arterial disease ever, n (\%) & 2084 (2\%) &
    249 (8\%)  &
    1835 (2\%) & 784 \\
    Pregnant ever, n (\%) & 84005 (90\%) &
    2760 (90\%)  &
    81245 (90\%) & 315 \\
    Angioplasty of coronary arteries ever, n (\%) & 1128 (1\%) &
    177 (6\%)  &
    951 (1\%) & 1509 \\
    Stroke ever, n (\%) & 1415 (2\%) &
    142 (5\%)  &
    1273 (1\%) & 56 \\
    \textbf{Health limits vigorous activities} &&&& \textbf{812}\\
    \hspace{5mm} Yes, limited a lot, n (\%) & 30022 (32\%) &

    1542 (51\%)  &
    28480 (32\%) & \\
    \hspace{5mm} Yes, limited a little, n (\%) & 41367 (45\%) &

    1162 (38\%)  &
    40205 (45\%) & \\
    \hspace{5mm} No, not limited at all, n (\%) & 21475 (23\%) &

    328 (11\%)  &
    21147 (24\%) & \\
    \hline
  \end{tabular}
    }
    \caption{Summary of Variables Influencing Treatment Assignment and Rule}
  \end{center}
\end{table}

\end{document}